\documentclass[twocolumn,preprintnumbers,amsmath,amssymb]{revtex4}
\usepackage{color}
\usepackage{graphics}
\usepackage{dcolumn}
\usepackage{bm}

\renewcommand\textfraction 0
\renewcommand\topfraction 1
\renewcommand\bottomfraction 1
\begin{document}
\title{Structural and magnetic properties of transition metal substituted ZnO} \author{S.
Kolesnik, B. Dabrowski,  and J. Mais } \affiliation{Department of Physics, Northern Illinois
University, DeKalb, IL 60115}
\date{\today}
\begin{abstract}
Structural and magnetic properties have been studied for polycrystalline Zn$_{1-x}$TM$_x$O, where
TM (transition metal ions) = Mn, Fe, and Co. No bulk ferromagnetism was observed for single-phase
materials, contrary to the existing theories. Single-phase samples demonstrate paramagnetic
Curie-Weiss behavior with antiferromagnetic interactions, similar to other diluted magnetic
semiconductors. Non-optimal synthesis conditions lead to formation of second phases that are
responsible for spin-glass behavior (ZnMnO$_3$ impurity for Zn$_{1-x}$Mn$_x$O (S. Kolesnik {\em et
al.}, J. Supercond.: Incorp. Novel Magn. {\bf 15}, 251 (2002))) or high-temperature ferromagnetic
ordering (Co metal for Zn$_{1-x}$Co$_x$O with the Curie temperature $T_C > 800$~K or
(Zn,Fe)$_{3}$O$_4$ for Zn$_{1-x}$Fe$_x$O with $T_C = 440$~K).

\vspace*{.5cm} Keywords: Diluted magnetic semiconductors, ZnO, synthesis, ferromagnetism\\
\end{abstract}


\maketitle

\section{Introduction}

A widegap II-VI semiconductor, ZnO attracts attention as a material with possible application in
optoelectronic devices such as solar cells, ultraviolet emitting diodes and transparent high-power
electronic devices. The discovery of ferromagnetism at temperatures above 100 K in the III-V
semiconductor Ga$_{1-x}$Mn$_x$As \cite{Ohno98} provided practical means to incorporate spin into
semiconductor electronics. Theoretical predictions of room temperature ferromagnetism in diluted
magnetic semiconductors \cite{Dietl00} recently focused attention on magnetic-ion-substituted ZnO
with a wurtzite structure similar to GaAs. According to these calculations, room temperature
ferromagnetism can exist in $p$-type doped Zn$_{1-x}$Mn$_x$O with 5\%Mn and $3.5 \cdot 10^{20}$
holes per cm$^3$. {\em Ab initio} band calculations \cite{Sato00a} predict stability of
ferromagnetism in $p$-type Zn$_{1-x}$Mn$_x$O, and antiferromagnetism in $n$-type Zn$_{1-x}$Mn$_x$O.
Similar calculations predict a ferromagnetic phase for both carrier-undoped and $n$-type ZnO
substituted with Fe, Co, or Ni. \cite{Sato00b} The parent compound ZnO without intentional carrier
doping shows $n$-type conduction related to oxygen vacancies and Zn interstitials.
\cite{Minegishi97} The introduction of $p$-type conduction in ZnO is difficult. Several methods,
including nitrogen doping \cite{Minegishi97,Look02} and Ga and N codoping
\cite{Yamamoto99,Joseph99}, have been reported. Successful preparation of $p$-type ZnO could allow
fabrication of transparent $p-n$ junctions and could enable synthesis of substituted ferromagnets.

Pulsed-laser deposited Zn$_{1-x}$Mn$_x$O thin films with up to 35\% Mn show spin-glass behavior.
\cite{Fukumura01}  We have previously shown for polycrystalline Zn$_{1-x}$Mn$_x$O that the
spin-glass behavior can be induced by high-pressure oxygen annealing and it is related to
precipitation of impurity phase ZnMnO$_3$. \cite{Kolesnik02} ZnO films doped with Co have been
reported to be ferromagnetic with Curie temperature of about 280 K. \cite{Ueda01} However, Kim {\em
et al.} pointed to the presence of Co clusters in their thin films as a reason for room temperature
ferromagnetism. \cite{Kim02} High-temperature ferromagnetism in thin films of Co-doped anatase
TiO$_2$ \cite{Matsumoto01} has also been explained by the fact that cobalt is not substitutional
for titanium in TiO$_2$. \cite{Stampe02} In this study, we investigate polycrystalline
Zn$_{1-x}$TM$_x$O, where TM = Mn, Fe, and Co. We show that no high-temperature ferromagnetic
ordering is present in single-phase samples. Ferromagnetism in these materials can be induced by
second phases that may appear after non-optimal synthesis conditions.

\section{Experimental methods}

The Zn$_{1-x}$TM$_x$O samples in this study were prepared using a standard solid-state reaction
technique. Mixtures of ZnO and MnO$_2$, Fe$_2$O$_3$, or Co$_3$O$_4$, for TM = Mn, Fe, and Co,
respectively, were fired in air several times at elevated temperatures 900, 1000, 1100, and
1200$^{\circ}$C with intermediate grindings. The samples were then annealed in an atmosphere of
various gases (Ar, H$_2$, 1\%H$_2$/Ar, and O$_2$ under high pressure of 600 bar). High pressure
oxygen annealing was intended to produce $p$-type doped ZnO materials. Magnetic ac susceptibility
and dc magnetization were measured using a Physical Property Measurement System and a Magnetic
Property Measurement System (both Quantum Design) at temperatures up to 400 and 800 K,
respectively. Energy dispersive x-ray spectroscopy (EDXS) analysis was performed by a Hitachi
S-4700-II scanning electron microscope. Typically, 15-20 EDXS spectra were collected for each
composition at various locations across the surface of a sintered pellet. Using the average Zn and
TM cation contents, which were obtained from the EDXS spectra we determined the effective
composition Zn$_{1-x_{\rm eff}}$TM$_{x_{\rm eff}}$O of the studied samples. X-ray diffraction
experiments have been performed using a Rigaku x-ray diffractometer. We have analyzed the x-ray
diffraction patterns using the Rietveld technique with the General Structure GSAS
code.\cite{GSAS85} The refinements were done using the wurtzite hexagonal space group P6$_{3}$mc
with (Zn,TM) atoms located at (2/3, 1/3, 0) and O atoms located at (2/3, 1/3, $u$). For both
single-phase and multi-phase samples (where impurity peaks were observed), this was the only phase
refined.

\section{Relationship between synthesis conditions and structural properties}

Single-phase samples can be synthesized in air for TM = Mn, and Co and in  1\%H$_2$/Ar for TM = Fe.
X-ray diffraction data show that the Zn$_{1-x}$TM$_x$O samples are single-phase with the wurtzite
structure up to a certain value of $x$. We estimated this value (the solubility limit) to be equal
to 0.1, 0.1, and 0.2 for TM = Mn, Fe, and Co, respectively. Above the solubility limit,
characteristic impurities can be observed by x-ray diffraction, namely ZnMn$_{2}$O$_4$, Fe$_{z}$O
or (Zn,Fe)$_{3}$O$_4$, and CoO for TM = Mn, Fe, and Co, respectively.

\begin{figure}[!]
 \resizebox{8.5cm}{!}{\includegraphics{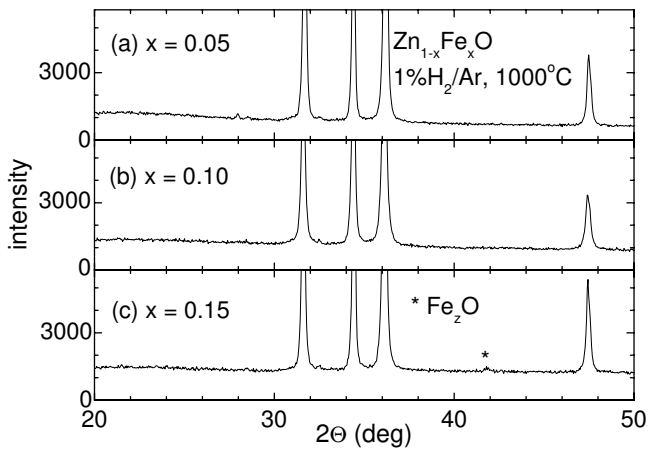}} \caption{\label{ZFxray} X-ray diffraction patterns
for Zn$_{1-x}$Fe$_{x}$O samples synthesized in 1\%H$_2$/Ar at 1000$^{\circ}$C.}
\end{figure}
In  Fig.~\ref{ZFxray} X-ray diffraction patterns are presented for Zn$_{1-x}$Fe$_{x}$O after
synthesis in 1\%H$_2$/Ar at 1000$^{\circ}$C. The samples with $x \leqslant 0.1$ are single-phase,
and the presence of the second phase Fe$_z$O can be observed for $x \geqslant 0.15$. W\"ustite
Fe$_z$O can typically be obtained in a slightly non-stoichiometric form with $z < 1$ depending on
synthesis conditions; however, due to a very small amount of this phase in the
Zn$_{0.85}$Fe$_{0.15}$O sample, we could not determine the value of $z$. Thus, we have separately
prepared w\"ustite Fe$_z$O by reduction of Fe$_{2}$O$_3$ in the same atmosphere, 1\%H$_2$/Ar at
1050$^{\circ}$C, and obtained the compound with the rock-salt structure and lattice constant $a =
4.3101$~\AA~that is consistent with $z = 0.95$. Stability of Zn$_{1-x}$Fe$_{x}$O and second phases
are sensitive to changes of the synthesis temperature. The annealing of these samples in
1\%H$_2$/Ar at 950$^{\circ}$C leads to precipitation of second phase (Zn,Fe)$_{3}$O$_4$, which
accounts for room temperature ferromagnetism in multi-phase Zn$_{1-x}$Fe$_{x}$O. Substitution of Fe
in ZnO is not possible in air or at high pressure of oxygen at any temperature.

\begin{figure}[!]
\resizebox{8.5cm}{!}{\includegraphics{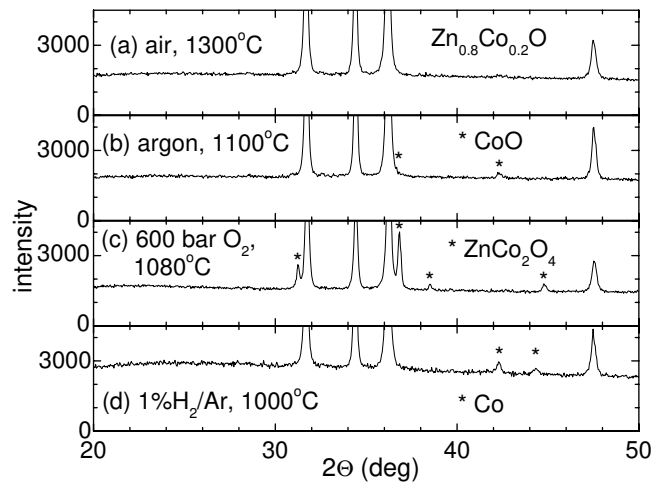}}
 \caption{\label{ZC02xray} X-ray diffraction patterns for
Zn$_{0.8}$Co$_{0.2}$O samples synthesized in air (a), argon-annealed (b), high-pressure
oxygen-annealed (c), and (d) slightly reduced in 1\%H$_2$/Ar at 1000$^{\circ}$C.}
\end{figure}
 Fig.~\ref{ZC02xray}
shows x-ray diffraction patterns for the Zn$_{0.8}$Co$_{0.2}$O sample obtained from several
annealing conditions. Various impurity peaks appear after annealing under reducing or strongly
oxidizing conditions. These peaks are marked with stars in Fig.~\ref{ZC02xray}. After argon
annealing at $T = 1100^{\circ}$C, rock-salt-like antiferromagnetic CoO impurity is present. A
slight reduction of Zn$_{0.8}$Co$_{0.2}$O in 1\%H$_2$/Ar at $1000^{\circ}$C leads to precipitation
of Co metal, which causes a ferromagnetic contribution to the measured ac susceptibility and dc
magnetization. High-pressure oxygen annealing at $T = 1080^{\circ}$C induces spinel-like
ZnCo$_2$O$_4$ impurity. High-pressure oxygen annealing was found, thus, to favor formation of
compounds other than Zn$_{1-x}$TM$_{x}$O, for which TM ions display oxidation states higher than
2+.

\begin{figure}[!]
 \resizebox{8.5cm}{!}{\includegraphics{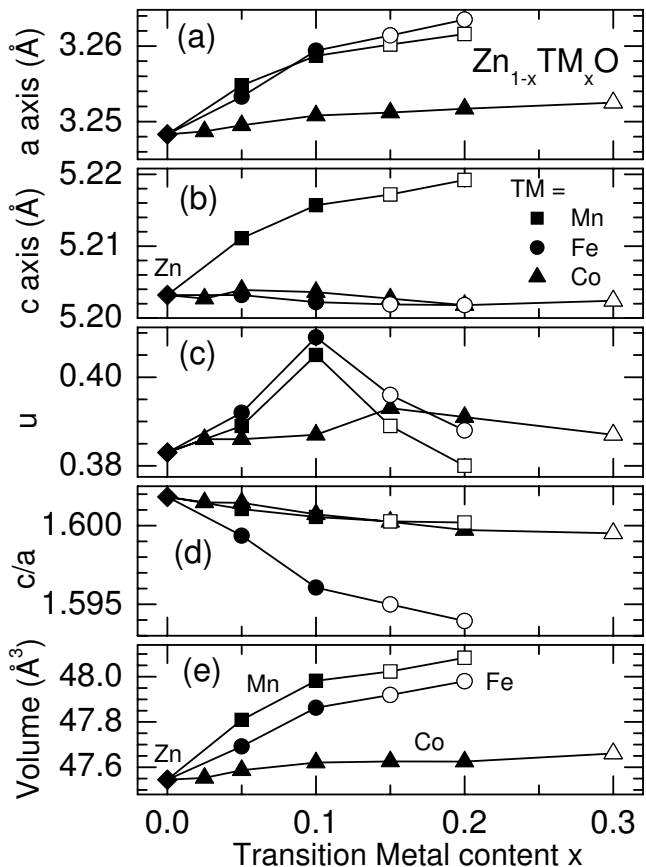}}
\caption{\label{ZTMlattpar} Lattice parameters a (a), c (b), u - parameter of the wurtzite
structure (c), c/a ratio (d), and unit cell volume (e) determined from the X-ray diffraction
patterns for Zn$_{1-x}$TM$_{x}$O samples. Full symbols denote single-phase samples, open symbols
denote multi-phase samples.}
\end{figure}
Structural parameters of Zn$_{1-x}$TM$_{x}$O obtained from the Rietveld refinements are presented
in Fig.~\ref{ZTMlattpar}. Open symbols in Fig.~\ref{ZTMlattpar} denote the multi-phase
compositions, i.e. the ones for which the presence of impurity peaks is observed in the X-ray
diffraction data. The unit cell volume changes almost linearly with substitution of TM for
single-phase compositions. This behavior is expected according to the difference between ionic
radii $r$ of tetrahedrally coordinated Zn$^{2+}$ $r = 0.60$~\AA~and other transition metal
TM$^{2+}$ ions ($r$ = 0.66, 0.63, and 0.58~\AA~ for TM = Mn, Fe, and Co, respectively).
\cite{Shannon76} The observed slight increase of the cell volume in case of TM = Co is inconsistent
with the smaller ionic radius of Co than Zn estimated by Shannon. \cite{Shannon76} This increase
can not be explained either by the presence of Co$^{3+}$ in Zn$_{1-x}$Co$_x$O as Co$^{3+}$ is
smaller than Co$^{2+}$. \cite{Shannon76} Apparently, the ionic size of tetrahedrally coordinated
Co$^{2+}$ ion is larger than that of Zn$^{2+}$. The $u$ parameter ($z$-coordinate of the oxygen
atoms) of the wurtzite structure of Zn$_{1-x}$TM$_x$O initially increases with increasing $x$,
reaches its maximum close to the solubility limit, and subsequently decreases. The $c/a$ ratio
[Fig. \ref{ZTMlattpar}(d)] decreases slightly at a similar rate for TM = Mn and Co, while it
decreases much more rapidly for TM = Fe. This effect shows that Fe-doped ZnO is more anisotropic
than ZnO doped with other transition metal ions. Ref. \cite{Schulz79} has shown a phenomenological
linear dependence $u\propto(a/c)^2$ for all undoped wurtzite-like compounds including ZnO. Our
observation for transition metal doped ZnO is qualitatively consistent with this tendency up to the
solubility limit. The coefficient of this linear dependence (varying from 9 to 34) is much higher
than the reported value of 1/3. \cite{Schulz79}

\begin{figure}[!]
 \resizebox{8.5cm}{!}{\includegraphics{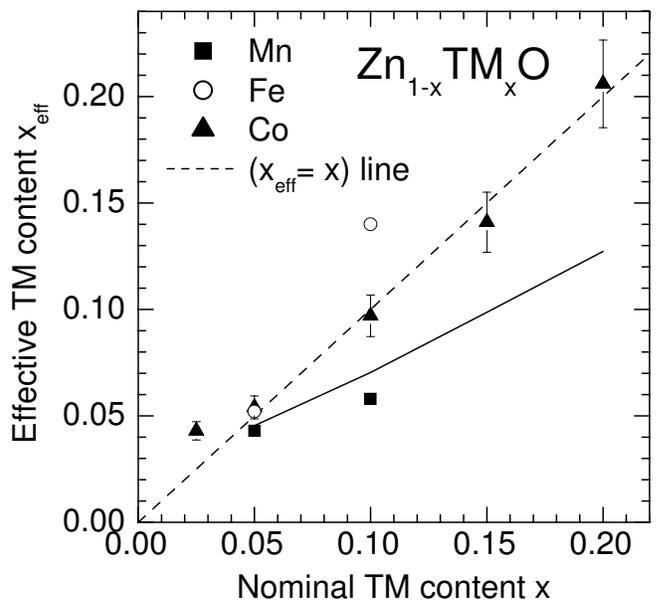}}
\caption{\label{TMcontents} The effective TM content in Zn$_{1-x}$TM$_{x}$O determined from the
EDXS analysis. Solid line shows $x_{\rm eff}$ for TM = Mn determined from the ac susceptibility
(see: Ref.~\cite{Kolesnik02}).}
\end{figure}
The effective TM content $x_{\rm eff}$ in Zn$_{1-x}$TM$_{x}$O determined from EDXS analysis is
presented in Fig.~\ref{TMcontents}. In our previous work~\cite{Kolesnik02} we have shown (based on
the assumption that the ac susceptibility reflects spin-only behavior) that $x_{\rm eff} < x$ for
TM = Mn (solid line in Fig.~\ref{TMcontents}.) Present EDXS data for TM = Mn confirm that result.
For TM = Co, we observe a good agreement between the effective and nominal TM contents, within the
experimental error. For TM = Fe, $x_{\rm eff} \simeq x$ for $x = 0.05$. However, for higher Fe
concentration, $x=0.1$, we observe $x_{\rm eff}\simeq 0.14$. This larger effective Fe content than
the starting content, $x$, could arise from Zn volatilization during synthesis under reducing
conditions in 1\%H$_2$/Ar. We did not include this composition for the analysis of the magnetic
properties of Zn$_{1-x}$Fe$_{x}$O.

\section{Magnetic properties}

\begin{figure}[!]
\resizebox{8.5cm}{!}{\includegraphics{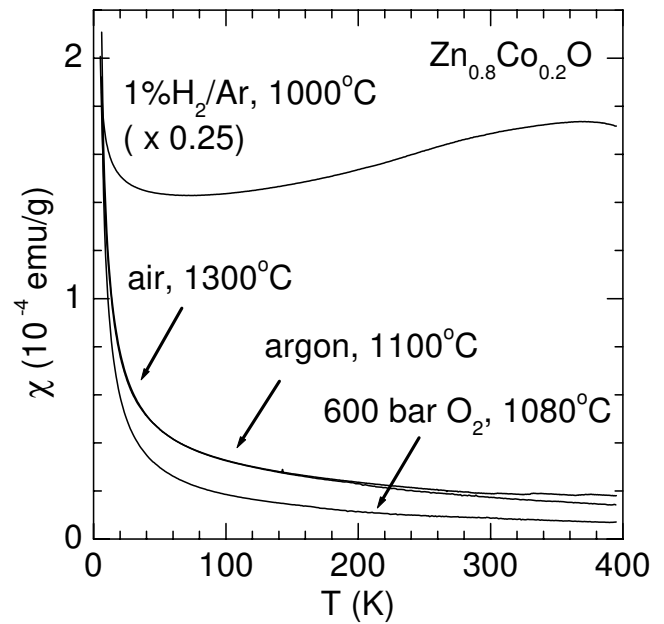}} \caption{\label{ZC02chi} Magnetic ac susceptibility
for Zn$_{0.8}$Co$_{0.2}$O samples synthesized under different conditions.}
\end{figure}
Magnetic susceptibility is presented in Fig. \ref{ZC02chi} for Zn$_{0.8}$Co$_{0.2}$O. Here and
throughout this paper, the diamagnetic ac susceptibility of $- 0.33 \cdot 10^{-6}$ emu/g for ZnO
\cite{Handbook00} was subtracted from the measured magnetic susceptibility. The ac susceptibility
of the TM-doped ZnO resembles Curie-Weiss behavior which is also characteristic for other
TM-containing semimagnetic semiconductors. \cite{Spalek86} The annealing in Ar under atmospheric
pressure (which is correlated to the appearance of antiferromagnetic CoO impurity [Fig.
\ref{ZC02xray}(b)]) does not significantly change the value of the ac susceptibility. A significant
decrease of the ac susceptibility can be observed after high-pressure oxygen annealing. After this
annealing, a spinel-like ZnCo$_{2}$O$_{4}$ second phase [Fig. \ref{ZC02xray}(c)] is present in the
sample. The most dramatic change of the magnetic properties of Zn$_{1-x}$Co$_{x}$O can be observed
after annealing in 1\%H$_2$/Ar at 1000$^{\circ}$C. The sample becomes ferromagnetic at a
temperature higher than 800~K. This ferromagnetic behavior is a result of the presence of Co-metal
impurity in the compound [Fig. \ref{ZC02xray}(d)].

\begin{figure}[!]
 \resizebox{8.5cm}{!}{\includegraphics{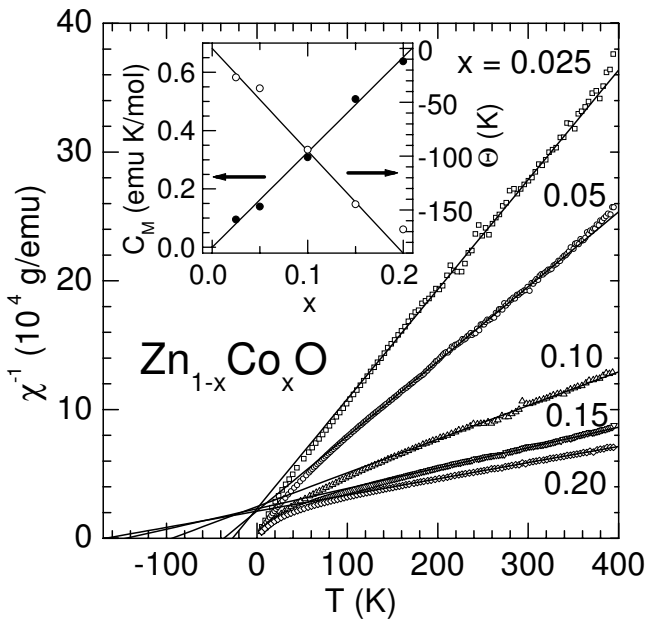}}
\caption{\label{ZCchi-1}   Inverse magnetic susceptibility for Zn$_{1-x}$Co$_x$O samples. The inset
shows molar Curie constants $C_M$ (full symbols) and Curie-Weiss temperatures $\Theta$ (open
symbols). }
\end{figure}
Inverse magnetic susceptibility for Zn$_{1-x}$Co$_x$O is shown in Fig. \ref{ZCchi-1}. The behavior
presented here is characteristic to all studied TM-doped ZnO samples. These results, for TM = Mn,
were presented elsewhere. \cite{Kolesnik02} We have analyzed ac susceptibility results within the
framework of the diluted Heisenberg antiferromagnet theory of Spa{\l}ek {\em et al.}
\cite{Spalek86} Although this theory was developed for larger group VI elements, Se, Te, and S, we
use it here also for the oxygen ion. At higher temperatures, the ac mass susceptibility can be
described by the formula
\begin{equation}
 \chi = \frac{C_M(x)}{T-\Theta(x)},
\end{equation}
where $\Theta(x) = \Theta_0 \cdot x$ is the Curie-Weiss temperature, $C_M(x) = C_0 \cdot x$ is the
molar Curie constant and $C_0$ is defined as
\begin{equation}
 C_0 = \frac{N(g_{\rm eff}\mu_B)^2S(S+1)}{3k_B\rho},
\end{equation}
$N$ is the number of cations per unit volume, $g_{\rm eff}$ is the effective gyromagnetic factor of
TM$^{2+}$ ion, $S$ is the spin, and $\rho$ is the mass density calculated from the lattice
parameters. The constant $\Theta_0$ is related to the exchange integral between the nearest TM
neighbors $J_1$,
\begin{equation}
\frac{2J_1}{k_B} = \frac{3\Theta_0}{zS(S+1)},
\end{equation}
where $z = 12$ is the number of nearest neighbors in the wurtzite structure of Zn$_{1-x}$TM$_x$O.

 A linear fit to the inverse susceptibility data intersects the $\chi^{-1} = 0$
 axis at a negative temperature. This result
indicates the presence of antiferromagnetic interactions in the Zn$_{1-x}$TM$_x$O samples.
 At lower temperatures, inverse ac susceptibility deviates from the linear dependence toward a temperature close to
zero. This is a result of additional antiferromagnetic interactions between the next nearest
neighbor TM ions.  \cite{Spalek86}. The inset to Fig. \ref{ZCchi-1} shows the molar Curie constants
and the Curie-Weiss temperatures for Zn$_{1-x}$Co$_x$O. From linear fits of $C_M$ and $\Theta$ as a
function of $x$, we have determined the parameters $C_0$ and $\Theta_0$ for the studied TM-doped
ZnO samples. The constant $C_0$ contains both the spin of the TM$^{2+}$ ions and the effective
$g_{\rm eff}$ factor. Mn$^{2+}$-substituted ZnO was already discussed in Ref. \cite{Kolesnik02} Due
to the half-filled $3d$ shell of Mn$^{2+}$, its orbital momentum $L=0$, the spin is $S=5/2$, and
the $g_{\rm eff}$ factor is very close to the free-spin value $g=2$. This is similar to other
Mn-based diluted magnetic semiconductors. For Co$^{2+}$, the splitting of the energy levels due to
the strong spin-orbit interaction results in an increase of the effective $g_{\rm eff}$ factor. The
values of $g_{\rm eff} = $2.24-2.31 were reported for Co$^{2+}$ in different diluted magnetic
semiconductors. \cite{Ham60,Koidl77,Lewicki89,Seong01,Alawadhi02} An enhanced $g_{\rm eff}$ factor
is also expected for Fe$^{2+}$ due to its large orbital momentum ($L=2, S=2$). \cite{Twardowski88}

 \begin{table}[htb] \caption{Material parameters determined from ac
susceptibility for Zn$_{1-x}$TM$_x$O.}
\begin{tabular}{cccccc}
\\
\hline \hline
TM ion & $\Theta_0$ (K) & $C_0$ (emu K/mol) & $2J_1/k_B$ (K) & spin $S$ & $g_{\rm eff}$\\
\hline
Mn &  -961$\pm$49 &  4.1$\pm$0.2 &  -27.5$\pm$1.4 &  5/2 & 2\\
Fe &  -926$\pm$26 &  6.0$\pm$0.1 &  -38.6$\pm$1.1 &  2 & 2.83\\
Co &  -951$\pm$110 &  3.3$\pm$0.1 &  -63.4$\pm$7.4 &  3/2 & 2.64\\
\hline \hline\\
 \end{tabular}
 \end{table}
In Table I, we compare the results determined from ac susceptibility for our samples with different
TM$^{2+}$. Here we assume the spin $S$ to be equal to the theoretically expected value for each
particular TM$^{2+}$ ion. Our values of $\Theta_0$ and $2J_1/k_B$ are similar to other
TM-containing semiconductors in case of TM = Mn \cite{Spalek86,Furdyna88} and TM = Co.
\cite{Alawadhi02} The effective $g_{\rm eff}$ factors for TM = Co and Fe are substantially larger
than the values from literature for other diluted magnetic semiconductors doped with respective
TM$^{2+}$ ions. The origin of this large $g_{\rm eff}$ factor  is not clear at the moment. The
presence of a small number of TM$^{3+}$ in our samples (not detectable by X-ray diffraction) is one
possibility. These ions can increase the effective spin above the assumed value for TM$^{2+}$. This
effect is especially crucial in the case of Fe substituted ZnO which we discuss below.

\begin{figure}[!]
 \resizebox{8.5cm}{!}{\includegraphics{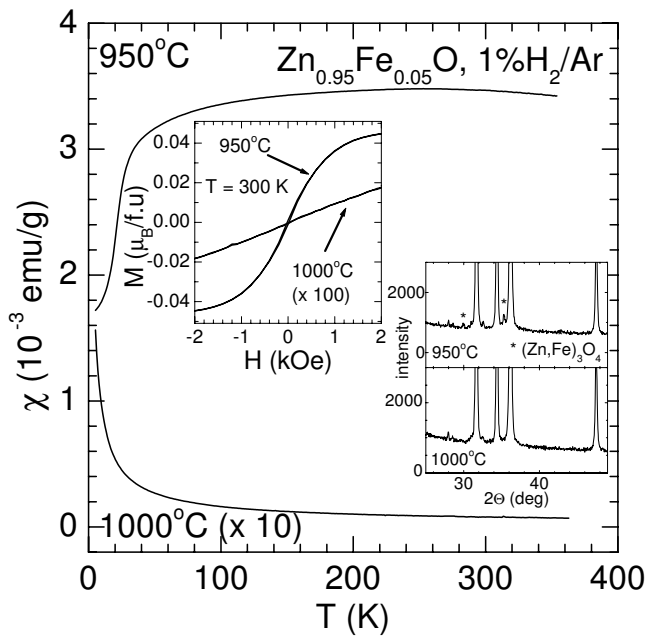}}
\caption{\label{ZF005} Magnetic ac susceptibility (main panel), dc magnetization at $T = 300$~K,
and X-ray diffraction patterns for Zn$_{0.95}$Fe$_{0.05}$O synthesized in 1\%H$_2$/Ar at
950$^{\circ}$C and 1000$^{\circ}$C.}
\end{figure}
The influence of synthesis conditions on the structural and magnetic properties of
Zn$_{0.95}$Fe$_{0.05}$O is illustrated in Fig.~\ref{ZF005}. The sample annealed at 950$^{\circ}$C
shows the presence of magnetite-like (Zn,Fe)$_3$O$_4$ impurity as displayed in the right inset. The
ac susceptibility of this sample is much larger than the paramagnetic susceptibility of pure
Zn$_{0.95}$Fe$_{0.05}$O (annealed at 1000$^{\circ}$C). The temperature dependence of ac
susceptibility for the sample containing (Zn,Fe)$_3$O$_4$ impurity is flat at higher temperatures,
which is related to a ferromagnetic order with the Curie temperature equal to 440 K. A drop of the
ac susceptibility around 20 K is probably related to the low temperature Verwey transformation,
similar to the one observed in stoichiometric Fe$_3$O$_4$ at 120 K. \cite{Bickford53} The left
inset to Fig.~\ref{ZF005} shows the magnetization curves at $T = 300$~K for both samples. The
(Zn,Fe)$_3$O$_4$-containing sample shows a narrow ferromagnetic hysteresis curve and the saturation
magnetization close to $0.05 \mu_B$ per formula unit, which is equal to $1 \mu_B$ per substituting
Fe ion. For ferrimagnetically coupled Fe ions in Fe$_3$O$_4$ one would expect the saturation
magnetization $1.33 \mu_B$ per Fe ion. This suggests that most of the Fe in the multi-phase
Zn$_{0.95}$Fe$_{0.05}$O sample is precipitated as a (Zn,Fe)$_3$O$_4$ impurity. The single-phase
Zn$_{0.95}$Fe$_{0.05}$O sample shows a linear paramagnetic dependence on the magnetic field with a
very small ferromagnetic hysteretic contribution. We were able to determine the magnitude of this
ferromagnetic contribution by the following numerical procedure. Linear fits of the $M(H)$ data
were performed in the high-field ranges 40 to 70 kOe and -40 to -70 kOe, where the ferromagnetic
contribution to the magnetization is expected to be fully saturated. The average of the two fitted
lines was subtracted from the measured $M(H)$ dependence. As a result, we obtained a small
ferromagnetic hysteresis curve. The saturation magnetization of this ferromagnetic contribution
gives an estimate for the upper limit of the ferromagnetic (Zn,Fe)$_3$O$_4$ impurity of 0.1\%.

\section{Summary}

In summary, by studying structural and magnetic properties of transition metal substituted
polycrystalline ZnO samples, we have found that no bulk ferromagnetism exists for single-phase
materials, contrary to the existing theories. Stoichiometric samples demonstrate paramagnetic
behavior with antiferromagnetic interactions similar to other semimagnetic semiconductors
substituted with transition metal ions. Selected synthesis conditions lead to impurity-induced
ferromagnetic order in Co- and Fe-doped ZnO. Attempts to induce $p$-type doping by high-pressure
oxygen synthesis lead to formation of secondary phases for which transition metals Mn, Fe, and Co
exist in high oxidation states.

\acknowledgments

This work was supported by NSF (DMR-0302617), the U.S. Department of Education, and the State of
Illinois under HECA. The EDXS analysis was performed in the Electron Microscopy Center, Argonne
National Laboratory, Argonne, IL.


\begin{thebibliography}{99}

\bibitem{Ohno98} H. Ohno, Science {\bf 281}, 951 (1998).
\bibitem{Dietl00} T. Dietl, H. Ohno, F. Matsukura, J. Cibert, and D. Ferrand, Science {\bf 287}, 1019 (2000).
\bibitem{Sato00a} K. Sato and H. Katayama-Yoshida, Jpn. J. Appl. Phys. {\bf 39}, L555, (2000).
\bibitem{Sato00b} K. Sato and H. Katayama-Yoshida, Jpn. J. Appl. Phys. {\bf 40}, L334, (2000).
\bibitem{Minegishi97} K. Minegishi, Y. Koiwai, Y. Kikuchi, K. Yano, M. Kasuga, and A. Shimizu, Jpn. J. Appl. Phys. {\bf 36}, L1453, (1997).
\bibitem{Look02} D. C. Look, D. C. Reynolds, C. W. Litton, R. L. Jones, D. B. Eason, and G. Cantwell, Appl. Phys. Lett. {\bf 81}, 1830, (2002).
\bibitem{Yamamoto99} T. Yamamoto et al., Jpn. J. Appl. Phys. {\bf 38}, L166, (1999).
\bibitem{Joseph99} M. Joseph, H. Tabata, and T. Kawai, Jpn. J. Appl. Phys. {\bf 38}, L1205, (1999).
\bibitem{Fukumura01} T. Fukumura, Z. Jin, M. Kawasaki, T. Shono, T. Hasegawa, S. Koshihara, and H. Koinuma, Appl. Phys. Lett. {\bf 78}, 958 (2001).
\bibitem{Kolesnik02} S. Kolesnik, B. Dabrowski, and J. Mais, J. Supercond.: Incorp. Novel Magn. {\bf 15}, 251 (2002).
\bibitem{Ueda01} K. Ueda, H. Tabata, and T. Kawai , Appl. Phys. Lett. {\bf 79}, 988 (2001).
\bibitem{Kim02} J. H. Kim, H. Kim, D. Kim, Y. E. Ihm, and W. K. Choo, J. Appl. Phys. {\bf 92}, 6066 (2002).
\bibitem{Matsumoto01} Y. Matsumoto, M. Murakami, T. Shono, T. Hasegawa, T. Fukumura, M. Kawasaki,
P. Ahmet, T. Chikyow, S. Koshihara, and H. Koinuma, Science {\bf 291}, 854 (2001).
\bibitem{Stampe02} P. A. Stampe, R. J. Kennedy, Y. Xin, and J. S. Parker, J. Appl. Phys. {\bf 92}, 7114 (2002).
\bibitem{GSAS85} A. C. Larson and R. B. von Dreele, {\em General Structure Analysis System}, University
of California, 1985-1990.
\bibitem{Shannon76} R. D. Shannon, Acta Cryst. A {\bf 32}, 751 (1976).
\bibitem{Schulz79} H. Schulz and K. H. Thiemann, Solid State Commun. {\bf 32}, 783 (1979).
\bibitem{Handbook00} {\em CRC Handbook of Chemistry and Physics}, 80th edn. 1999-2000, p. 4-136.
(CRC Press LLC, Boca Raton, London, New York, Washington, D. C. 1999).
\bibitem{Spalek86} J. Spa{\l}ek, A. Lewicki, Z. Tarnawski, J. K. Furdyna, R. R. Ga{\l}azka, and Z. Obuszko, Phys. Rev. B {\bf 33}, 3407 (1986).
\bibitem{Twardowski87} A. Twardowski, H. J. M. Swagten, W. J. M. de Jonge, and M. Demianiuk, Phys. Rev. B {\bf 36}, 7013 (1987).
\bibitem{Furdyna88} J. K. Furdyna, N. Samarth, R. B. Frankel, and J Spa{\l}ek, Phys. Rev. B {\bf 37}, 3707 (1988).
\bibitem{Ham60} F. S. Ham, G. W. Ludwig, G. D. Watkins, and H. H. Woodbury, Phys. Rev. {\bf 5}, 468 (1965).
\bibitem{Koidl77} P. Koidl, Phys. Rev. B {\bf 15}, 2493 (1977).
\bibitem{Lewicki89} A. Lewicki, A. I. Schindler, J. K. Furdyna, and W. Giriat, Phys. Rev. B {\bf 40}, 2379 (1989).
\bibitem{Seong01} H. J. Seong, H. Alawadhi, I. Miotkowski, A. K. Ramdas, and S. Miotkowska, Phys. Rev. B {\bf 63}, 125208 (2001).
\bibitem{Alawadhi02} H. Alawadhi, I. Miotkowski, A. Lewicki, A. K. Ramdas, S. Miotkowska, and M. McElfresh, J. Phys.: Condensed Matter {\bf 14}, 4611 (2002).
\bibitem{Twardowski88} A. Twardowski, A. Lewicki, M. Arciszewska, W. J. M. de Jonge, H. J. M. Swagten, and M. Demianiuk, Phys. Rev. B {\bf 38}, 10~749 (1988).
\bibitem{Bickford53} L. R. Bickford, Phys. Rev. {\bf 25}, 75 (1953).
\end{thebibliography}
 \end{document}